\documentclass[twocolumn]{webofc}

\usepackage[varg]{txfonts}   
\usepackage{hyperref}
\usepackage{url}
\usepackage{titlesec}
\hypersetup{colorlinks=true,citecolor=blue,urlcolor=blue,linkcolor=blue}
\usepackage{caption}
\begin{document}
\title{Electric dipole excitations near the neutron separation energies in $^{96}$Mo}
%
%

\author{\firstname{Eun Jin} \lastname{In}\inst{1}\fnsep\thanks{\email{in2@llnl.gov}} \and
        \firstname{Emanuel} \lastname{Chimanski}\inst{2} \and
        \firstname{Jutta} \lastname{Escher}\inst{1} \and
        \firstname{Sophie} \lastname{P{\'e}ru}\inst{3} \and
        \firstname{Aaina} \lastname{Thapa}\inst{1} \and
        \firstname{Walid} \lastname{Younes}\inst{1}
}

\institute{Lawrence Livermore National Laboratory, Livermore, CA, USA $^{1}$
\and
           Brookhaven National Laboratory, Upton, NY, USA $^{2}$
\and
           CEA, DAM, DIF, Arpajon, France $^{3}$
          }

\abstract{
Electric dipole strength near the neutron separation energy significantly impacts nuclear structure properties and astrophysical scenarios. These excitations are complex in nature and may involve the so-called pygmy dipole resonance (PDR).
Transition densities play a crucial role in understanding the nature of nuclear excited states, including collective excitations, as well as in constructing transition potentials in DWBA or coupled-channels equations.
In this work, we focus on electric dipole excitations in spherical molybdenum isotopes, particularly $^{96}$Mo, employing fully consistent Hartree-Fock-Bogoliubov (HFB) and Quasiparticle Random Phase Approximation (QRPA) methods. We analyze the dipole strength near the neutron separation energy, which represents the threshold for neutron capture processes, and examine the isospin characteristics of PDR states through transition density calculations.
Examination of proton and neutron transition densities reveals distinctive features of each dipole state, indicating their isoscalar and isovector nature. We observe that the primary component in the enhanced low-energy region exhibits isovector character. The PDR displays a mixture of isoscalar and isovector nature, distinguishing it from the isovector giant dipole resonance (IVGDR).
These findings lay the groundwork for future investigations into the role of transition densities in reaction models and for their application to inelastic scattering calculations.
}
\maketitle

\section{Introduction}
\label{intro}
The origin and nature of the pygmy dipole resonance (PDR), a small enhancement in the dipole response function in the lower tail of the GDR near the neutron separation energy, has been extensively studied in recent years.
The PDR has been observed in both stable and unstable nuclei.
In neutron-rich nuclei, the radial distribution of excess neutrons extends further than that of protons, leading to the formation of a neutron skin. The oscillation of this skin is linked to the PDR. Microscopic structure models, especially self-consistent mean-field based models, have greatly advanced our understanding of this mode~\cite{Paar2007,Lanza2023}. Transition densities reveal that the PDR displays unique characteristics, notably in-phase oscillations within the nucleus and neutron-dominated oscillations at the surface, distinguishing it from the isovector giant dipole resonance (IVGDR). Indeed, the PDR region can encompass both isoscalar and isovector characteristics, as well as the pygmy mode within these dynamic processes. Despite these advancements, the nature of the pygmy mode requires further investigation.

Besides its importance from a nuclear structure perspective, the PDR plays a crucial role in the nucleosynthesis of heavy elements in nuclear astrophysics. The PDR enhances neutron capture cross sections, such as in reactions like $^{205}$Pb(n,$\gamma$)$^{206}$Pb which are significant in the astrophysical s-process~\cite{Tonchev2017}. Therefore, a detailed analysis of the dipole strength near the neutron separation energy is valuable not only for understanding the nuclear structure of the PDR but also for its role in the synthesis of heavy nuclei. 

In this proceeding, we present predictions for electric dipole excitations in spherical Mo isotopes, particularly $^{96}$Mo, which are of considerable interest due to their involvement in various astrophysical processes such as the $r$-, $s$-, and $p$-processes~\cite{Wieser2007,Stephan2019}. Our focus is on the energy region near the neutron separation energy, as it represents the threshold for neutron capture processes to occur. Through the analysis of the transition densities, we show the nature of excited states in the PDR region and discuss the isospin characteristics of pygmy states. In future work, the microscopically calculated transition densities will be used in folding calculations to produce input for proton- and neutron-induced scattering calculations.

\section{Transition densities as a tool to investigate nuclear structure and predict scattering}
\label{tool}
Information on wave functions and transition densities from many-body nuclear structure models provides critical insights into the nature of nuclear excited states, including collective excitations. Transition densities, in particular, are crucial for probing the isoscalar, isovector, or exotic character of these states, as they reveal changes in the nucleon density distribution within a nucleus during excitation. 

While phenomenological potentials fitted to experimental data have successfully described scattering processes for stable nuclei, they may fall short for unstable nuclei due to scarce experimental data. In response, microscopic approaches based on nuclear structure information, independent of experimental data, have been developed, with the Jeukenne-Lejeune-Mahaux (JLM) approach being widely adopted~\cite{Bauge1998,Bauge2001,Dupuis2019,Aaina2024}. 
Transition densities serve as essential ingredients for constructing transition potentials used in Distorted Wave Born Approximation (DWBA) or coupled-channels calculations.
The application of these transition potentials enables more accurate nuclear data evaluations for neutron-induced reactions~\cite{Dupuis2019} and enables surrogate experiments aimed at constraining neutron capture cross sections~\cite{Jutta2012,Andrew2019,Perez2020}.

We compute transition matrix elements from the ground state to excited states, from which we derive the dipole response functions and radial transition densities. We obtain the state information by solving the nuclear many-body problems, using the Hartree-Fock-Bogoliugov (HFB) and the Quasiparticle Random Phase Approximation (QRPA) in a fully consistent manner, employing the same Gogny D1M finite-range effective interaction. The HFB calculations provide the ground state energy and its wave function. We then use the QRPA to construct the vibrational excited spectra of the nucleus from correlated two-quasiparticle configurations built upon the HFB solutions.

The electric dipole response from the ground state to excited states is computed for the electromagnetic and isoscalar (IS) dipole operators, defined as:
\begin{eqnarray}
\hat{O}_{E1} = \frac{eZ}{A} \sum_{n=1}^{N} r_n Y_{1m}(\hat{r}_n) - \frac{eN}{A} \sum_{p=1}^{Z} r_p Y_{1m}(\hat{r}_p), \\
\hat{O}_{IS} = \sum_{i=1}^{A} \left(r_i^3 - \frac{5}{3} \left<r^2\right> r_i \right) Y_{1m}(\hat{r}_i),
\end{eqnarray}
where the $p$ and $n$ indices denote protons and neutrons, respectively, and $i$ runs over all nucleons.
The $\hat{O}_{E1}$ and $\hat{O}_{IS}$ operators include corrections to restore translational invariance.
Effective charges for protons ($eN/A$) and neutrons ($-eZ/A$) are introduced to remove the center of mass motion, which is a spurious translational mode. 
The isoscalar dipole operator $\hat{O}_{IS}$ has been modified by including the term with $5<r^2>/3$ to eliminate the center of mass component, as the leading-order term corresponds to a spurious translational motion.

\section{Results}

\subsection{HFB ground state densities and QRPA dipole response}

\begin{figure}
\centering
\includegraphics[width=0.35\textwidth,clip]{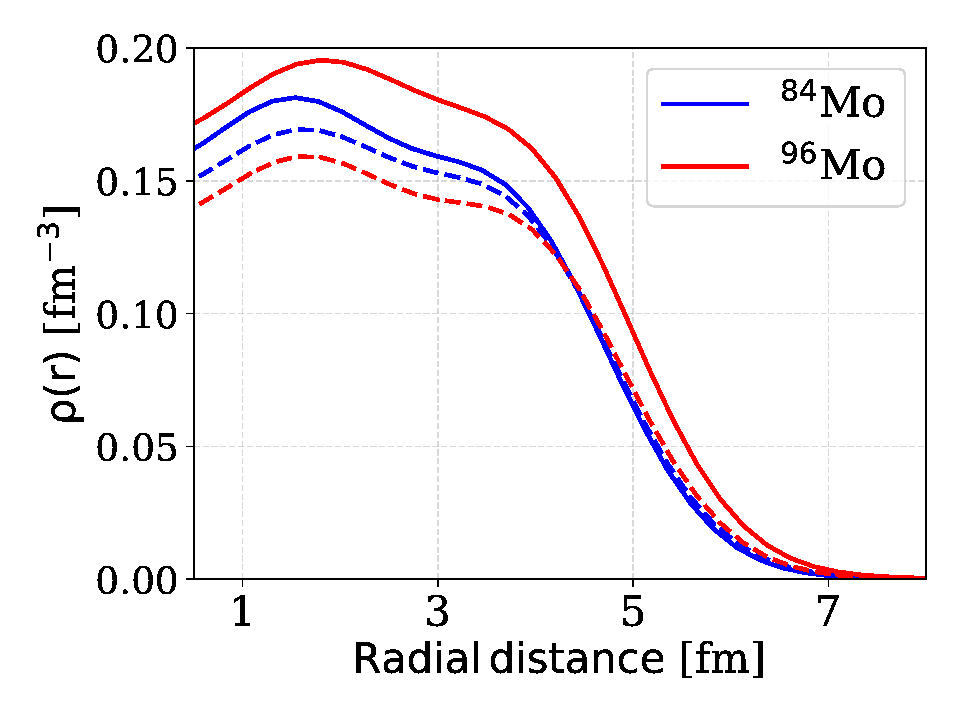}
\captionsetup{skip=0pt}
\caption{The neutron and proton ground state densities as a function of radial distance $r$ for $^{84}$Mo (blue) and $^{96}$Mo (red) isotopes are presented by solid and dashed lines, respectively.}
\label{gsdensity}       
\end{figure}

Figure~\ref{gsdensity} shows the proton and neutron spatial densities of the ground states for $^{84,96}$Mo isotopes. 
Our focus centers on the surface region, at the ground state root-mean-square radius for $^{96}$Mo of approxinately 4.3 fm, and beyond.
We observe significant excess in the neutron density in this space range, indicating the formation of a neutron skin. $^{84}$Mo, on the other hand, does not exhibit such skin.

\begin{figure}
\centering
\includegraphics[width=0.26\textwidth]{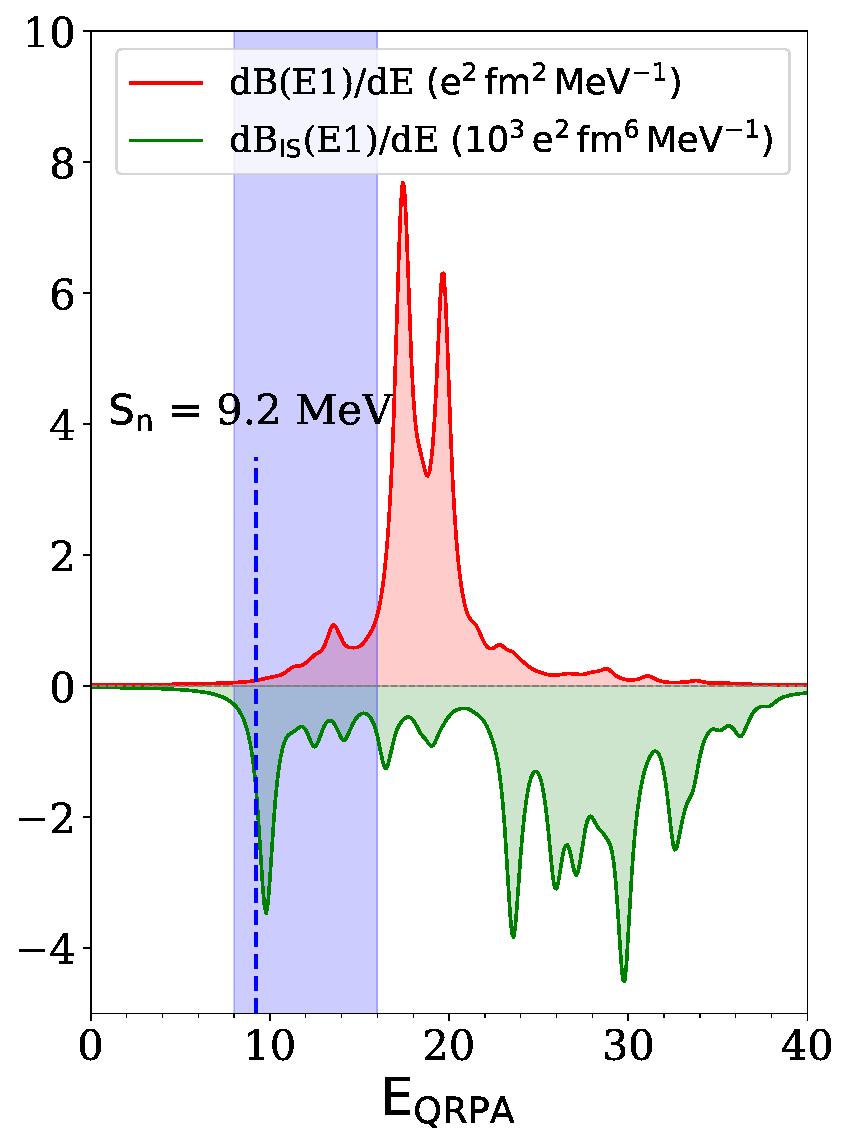}
\captionsetup{skip=0pt}
\caption{The electric dipole $(dB(E1)/dE)$ and isoscalar $(dB_{IS}(E1)/dE)$ response functions for $^{96}$Mo as a function of the QRPA excitation energy. The dashed vertical line represent the neutron separation energy for $^{96}$Mo. The PDR region is highlighted by a vertical band.}
\label{response-96mo}       
\end{figure}

Figure~\ref{response-96mo} shows the effects of neutron excess to the electric dipole response. The figure illustrates isovector and isoscalar dipole strength as continuous curves, obtained by folding the discrete spectrum with Lorentzian functions with a width $\Gamma = 1$ MeV. We observe a small enhancement in the dipole strength distribution near the neutron separation energy. 
This enhancement is characteristic of the PDR and serve as a potential singular piece of experimental evidence supporting its presence in $^{96}$Mo.

\subsection{Features of radial transition densities}

\begin{figure}
\centering
\includegraphics[width=0.45\textwidth]{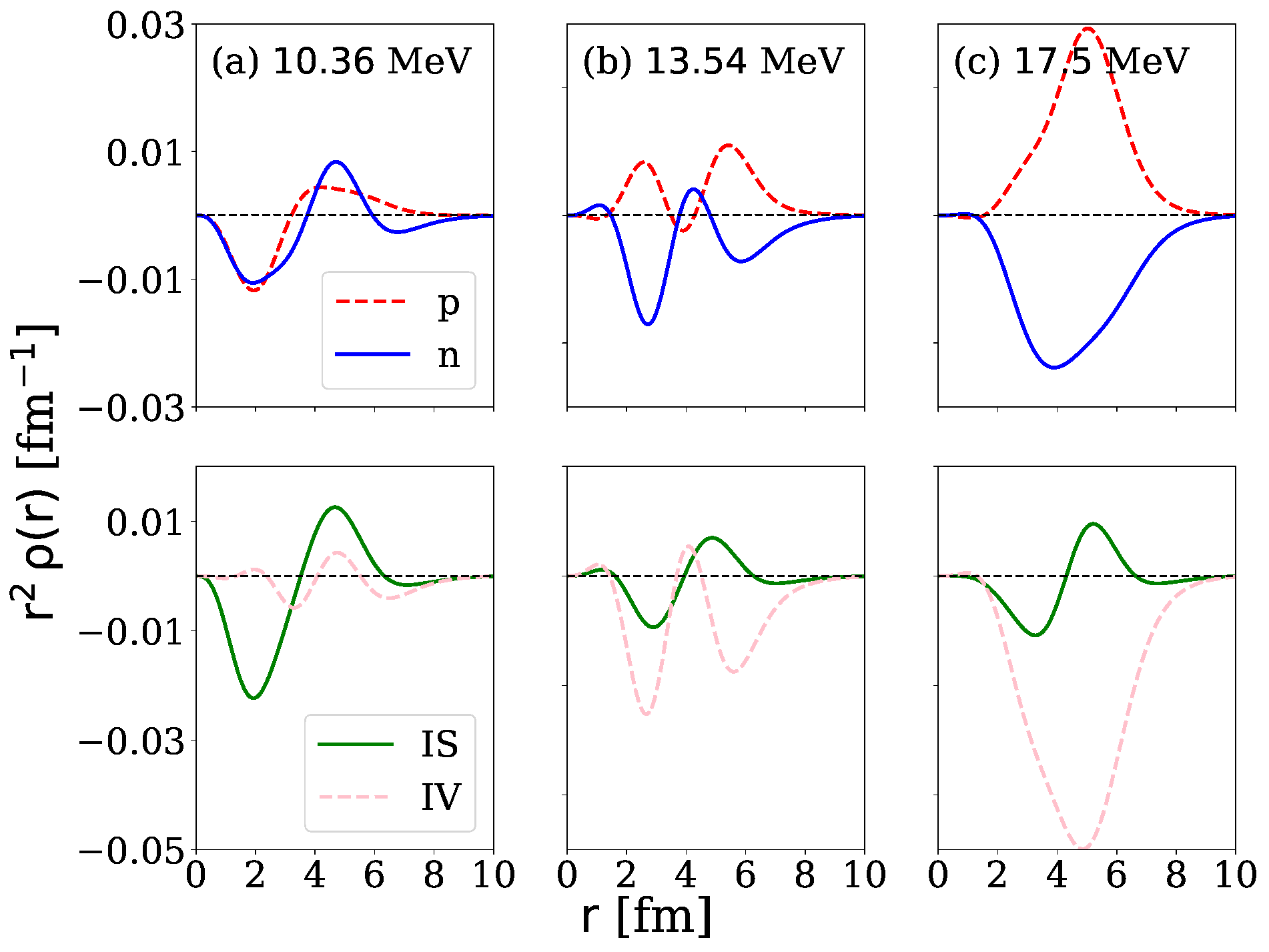}
\captionsetup{skip=0pt}
\caption{Radial transition densities for $^{96}$Mo isotope are shown for three cases: (a) the neutron PDR candidate, (b) the major peak of the enhancement region, and (c) the major GDR peak with the largest $B(E1)$. 
}
\label{nPDR_radial_td} 
\end{figure}

Figures~\ref{nPDR_radial_td} displays the radial transition densities of representative states in $^{96}$Mo, including two states in the potential PDR region (a+b), and the major GDR state with the largest $B(E1)$ values (c). 
In panel (a), the neutron and proton components of the transition densities oscillate in phase within the interior region. 
In contrast, for large $r$ ($r$ $\geq$ 7 fm), the proton contribution vanishes, and only the neutron component significantly contributes to both isoscalar and isovector transition densities, with nearly equal magnitude.
This behavior conforms to what one expects for the PDR~\cite{Vretenar2001,Tsoneva2008,Colo2009}. On the other hand, the largest peak in the potential PDR region, at 13.5 MeV in $^{96}$Mo, exhibits an isovector nature similar to that of the IVGDR. 
Further investigation is needed to better understand this enhancement near the neutron separation energy.

\section{Conclusion and outlook}

We have investigated the electric dipole response of spherical molybdenum isotopes, focusing on $^{96}$Mo, to explore the existence and nature of pygmy dipole states in molybdenum nuclei. Our analysis is based on fully consistent Hartree-Fock-Bogoliubov (HFB) and Quasiparticle Random Phase Approximation (QRPA) calculations using the Gogny D1M effective interaction.
We observed an enhancement in dipole strength near the neutron separation energies, which correlates with neutron excess. Our examination of proton and neutron transition densities reveals distinctive features for each dipole state, including their isoscalar and isovector characteristics. We see that the dominant components in the enhanced low-energy region exhibit isovector nature. A PDR state was also identified. It displays a mixture of isoscalar and isovector characteristics, distinguishing it from the isovector giant dipole resonance (IVGDR).
In future work, the calculated transition densities will serve as inputs for inelastic scattering calculations.

\section*{Acknowledgments}
This work was performed under the auspices of the U.S. Department of Energy by Lawrence Livermore National Laboratory under Contract DE-AC52-07NA27344 with partial support from LDRD Projects 22-LW-029 and 19-ERD-017.

%
\bibliography{myBib.bib} 
%
%
%
%

\end{document}